# Development of a Scaled Down Horizontal Axis Wind Turbine for Wind Tunnel Tests

Bhaskar Upadhyay Aryal[1], Bibek Bhurtel[1], Amit Sharma Bhandari[1], Sailesh Chitrakar[2] and Binaya Baidar[1]

[1]Department of Mechanical Engineering, School of Engineering, Kathmandu University
Turbine Testing Lab, Dhulikhel, Nepal, bhaskar.aryal@student.ku.edu.np, bibekbhurtel2048@gmail.com, amitsharmabhandari@gmail.com, binayabaidar@ku.edu.np
[2] Department of Energy and Process Engineering, Norwegian University of Science and Technology Trondheim, Norway, sailesh.chitrakar@ntnu.no

## Abstract

This paper presents a method to design and develop scaled down models of Horizontal Axis Wind Turbine (HAWT) to test in a limited sized wind tunnel. The mathematical design of a HAWT is carried out at a rated speed of 5 m/s, by using Qblade, which is based on BEM theory. The 2.5 m long blade is scaled to 0.15 m with geometric, kinematic and dynamic similarity conditions. This model was tested in an existing open circuit wind tunnel with $0.38 \times 0.38$ m$^2$ test section at Fluid Lab of Kathmandu University. Physical model of the turbine is developed using additive manufacturing method. For experiments conducted at pitch angles 0° and 5°, the turbine's response is found to be better at the latter angle with maximum coefficient of performance 0.32. This paper also presents graphical analysis of experimental results along with computational analysis for model validation. The method described in this paper can be implemented for conducting a cost-effective experimental validation of wind turbines designed for laboratory or industrial purposes.

## 1. Introduction

Wind turbines extract kinetic energy from the wind and convert it into other useful forms such as mechanical and electrical energy. With increasing energy demand and pollution problems associated with fossil fuels, the establishment of wind energy as a sustainable alternative source is imperative. No significant contribution to carbon emission to the atmosphere and its abundant availability make it one of the most viable energy alternatives [1].

Generation of wind power depends upon the interaction of the wind with the turbine rotor i.e. the aerodynamic forces generated by the wind determine the major aspects of wind turbine performance. Wind tunnel tests are conducted to study the aerodynamic behaviour of a wind turbine and extract valuable information to improve and optimize rotor design for harnessing more power from the available wind. Experimental approach is also necessary to validate the computational studies conducted in most of the researches these days. However, for large size blades, the size of a conventional wind tunnel is inappropriate to conduct experiments. These performance evaluation tests are conducted by scaling down turbine in accordance to the available wind tunnel using similarity laws [2-3].

BEM is widely used to predict the efficiency and performance of wind turbine. The design and analysis software QBlade implements BEM model which is cost effective and has significantly less computational time. More to that, repeated tests with small changes in design is possible and the user can compare two or more design details [4]. The number of blades in a horizontal axis wind turbine can vary. Usually, they have three blades but turbines with two and even one blade are also available. The polar moment of inertia for a three bladed rotor is independent of the azimuthal position of rotor and constant in respect to yawing. This in turn assures the smooth operation of the rotor in case of three-blade rotor. The difference in moment of inertia when the blades are at vertical and horizontal positions in case of two-blade turbine causes imbalance. More than three blades are not preferred because of the additional cost associated [5]. The power in the wind is limited. So, greater number of blades means extraction of less power by an individual blade. Rotor solidity is the total blade area as a fraction of the (annular disc) swept area, and there is an optimum solidity value for a given tip speed. To maintain this optimum value of rotor solidity, the higher the number of blades, the narrower each ought to be. For the air to pass through easily, the blades must be thin in relation to their width. So, the limited solidity also limits the thickness of the blades.

Burdett and Treuren tested different turbine models of 0.5, 0.4 and 0.3 m diameter with tunnel blockage up to 52.8 % to demonstrate scaling using Reynolds number matching. Besides geometric similarity and tip speed ratio matching, Reynolds number matching is also necessary for wind tunnel tests involving Reynolds number less than 500,000 [2]. Brusca et al. [6] determined the performance of wind turbine as a function of geometric similarity coefficient. They state that by expressing all the results as function of geometric similarity coefficient, it is possible to identify a group of wind turbines with same high performance but different geometrically.

Researches involving wind turbine blade's experimental investigation of aerodynamic characteristics through wind tunnel testing, are limited in public domain. Majority of studies conducted use Computational Fluid Dynamics and numerical codes based on BEM theory [7]. However, in this study the horizontal axis wind turbine blade designed on the principles of BEM theory by Aryal et.al [8] is analysed using computational method and validated with experimental aerodynamic investigations through wind tunnel test. To develop the experimental setup, turbine model is constructed using rapid prototype technology. The turbine is tested under different potential configurations (of pitch angle and placement on the test section) in order to enhance its performance; and, power coefficients are compared in each case. The characteristics of power, torque and efficiency were studied in all different configurations.

## 2. Turbine Design

### 2.1 Wind Energy

When mass of air passes through a wind turbine it loses its kinetic energy and the flow gradually slows down. The turbine transforms the kinetic energy in the wind into mechanical energy and then into electric energy. To obtain maximum power from the passing wind, the wind speed should be reduced to zero after passing through the turbine [9]. The energy of certain mass of air moving through a given surface area is given by eq. (1).

$$P = \frac{1}{2}\rho A v^3 \qquad (1)$$

Where, $P$– the maximum available energy in the wind, $\rho$ – air density, $A$ – surface area and $v$ – wind speed.

The relation shows that the power increases with the cube of wind speed but only linearly with density and area. As it is practically impossible to reduce the wind speed to zero, a relation describing the theoretical maximum efficiency was established by Albert Betz; denoted as Betz's limit and equal to 0.593 independent of the design of wind turbine [9-10]. The power coefficient $C_P$ that characterizes turbine performance is expressed by eq. (2).

$$C_P = \frac{P_R}{\frac{1}{2}\rho A v^3} \qquad (2)$$

The power produced by the turbine is further reduced by mechanical and electrical losses.

### 2.2 Blade Element Momentum Theory

The classical Blade Element Momentum (BEM) Theory is the combination of momentum theory also called disk actuator theory with blade element theory. The disk actuator theory is a mathematical model of an ideal actuator disk. The forces and flow conditions on a wind turbine blade can be derived considering the conservation of momentum equation. From this momentum theory, relations [5] can be derived which define thrust and torque as functions of axial and angular induction factors on the annular section (see Fig. 1.) of a rotor. Calculation of angle of attack along with lift and drag coefficients of the blade profile is possible with the wind speed of every section, computed using momentum theory. Blade Element Theory is used to express forces on the blade as a function of lift and drag coefficient and angle of attack. Forces are determined on each of the discretized individual element and integrated later along the whole span of the blade. Figure 1. shows the aerodynamic forces and angles in a section of the blade. Radial position, chord, twist and length define each cross section of the blade. This theory assumes that no aerodynamic interaction takes place between the blade elements [4, 11].

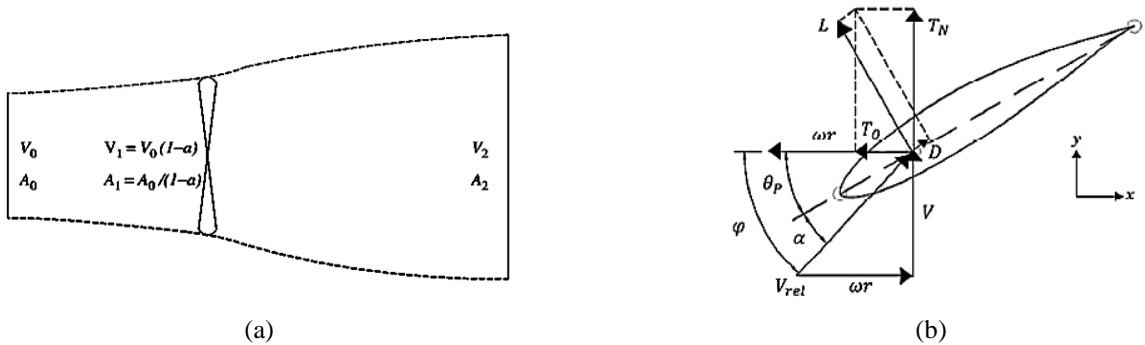

(a) (b)

**Fig. 1** (a) Annular areas ($A_0$, $A_1$, $A_2$) of actuator disk model [11] and (b) aerodynamic forces and angles of a blade section [13]

For further details on BEM theory, blade element equations and momentum equations can be referred to Jamieson [12].

### 2.3 Influential Design Parameters

Rotor blades convert the kinetic energy in the wind to useful mechanical and electrical energy. The shape of the blade is influenced primarily by different aerodynamic and structural considerations among a lot of other factors like blade material, manufacturing, recyclability, noise reduction etc. Rated power and wind speed, tip speed ratio, number of blades, rotor solidity and orientation (upwind or downwind), airfoil are some of the aerodynamic factors affecting blade design. Rated power is directly related to wind speed, sweep area and hence length of the blade [5]. The choice of different airfoils along the span of the blade is important for both aerodynamic performance and structural rigidity. Foils near the blade root should provide strength without hampering the blade performance. The blade is designed using circular foils in the root section and NACA (National Advisory Committee for Aeronautics) foils in the tip region. Full scale design of blade has a length of 2.5 meters and the rated speed is 5 m/s.

Figure 2 shows the blade modeled using SolidWorks with design based on BEM theory (a) and physical model of wind turbine test stand.

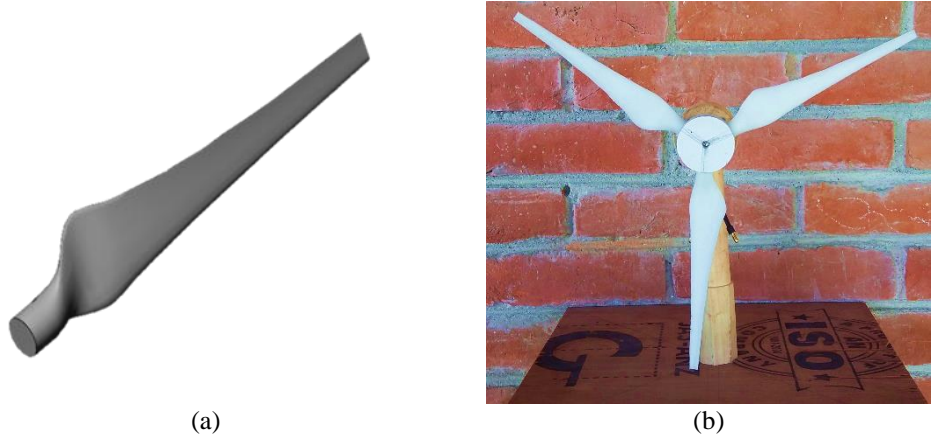

(a)          (b)

**Fig. 2** (a) Blade model drawn in SolidWorks and (b) Assembled scaled tri-blade turbine model

The ratio of speed of the tip of a turbine blade to the undisturbed freestream wind velocity is tip speed ratio (λ). It is a dimensionless number defining how many more times the tip speed is than the wind speed. The chosen design tip speed ratio should maximize the coefficient of power. Low solidity is obtained with higher tip speed ratio which in turn reduces the total blade area resulting in lighter blades; but it increases the aerodynamic noise, vibration and higher stress. For modern wind turbines with three blades, TSR usually varies from 5 to 7. The tip speed ratio has effect on twist and chord of blade (see eq. 5,6 and7) [5,14]. For rotational speed ($\omega$), Blade radius $R$ and freestream wind velocity $v$, TSR is given by eq. (3).

$$\lambda = \frac{\omega R}{v} \qquad (3)$$

The angle between chord line of the airfoil and the direction of apparent wind – angle of attack ($\alpha$) has an optimum value for the design which maximizes the lift and hence the performance of the blade. The optimum value of $\alpha$ obtained from simulations using QBlade for this design is 5 degrees. Similarly, the (critical) angle of attack above which stall occurs is 18 degrees. Above stall angle, the airflow on the upper surface of the airfoil becomes less smooth resulting in boundary separation. The flow cannot attach itself from leading edge to trailing edge and flow separation occurs [15]. The aerodynamic force can be resolved into lift and drag components. These components generate normal force i.e. thrust and tangential force i.e. torque.

To define each section of the blade one of the important parameters is twist angle ($\beta$). Twist angle is based on aerodynamic behavior of that individual section at certain angle of attack. It is necessary for better starting torque and stronger blade root. The relative angle of wind ($\varphi$) is expressed by eq. (4) [5]

$$\varphi = tan^{-1} \frac{2}{3\lambda_r} \qquad (4)$$

where, $\lambda_r$ is local tip speed ratio at radius $r$ of the blade.

The chord length of airfoil in the blade varies as the blade is slender and tapered towards the tip region and circular in the root region. The tip portion of the blade has to process more wind whereas the root should be structurally integral, and hence they have different chord lengths and shapes. The chord distribution at any blade radius ($C(r)$) can be obtained according to eq. (5) given by Betz [4].

$$C(r) = \frac{16}{9} \frac{\pi R}{B C_L \lambda_r} \frac{1}{\sqrt{\left(\lambda_r \frac{r}{R}\right)^2 + \frac{4}{9}}} \qquad (5)$$

Where, $R$ – radius of blade, $B$ – number of blades, $C_L$ – coefficient of lift, $r$– local radius of blade, $\lambda_r$ – local tip speed ratio at radius $r$.

From eq. (5), it can be inferred that shape of the blade should be tapered as chord length ($C$) is inversely proportional to radius ($r$), also fewer blades mean wider blades.

Blade pitch angle (angle between tip chord and rotor plane) of a wind turbine rotor affects its aerodynamic performance. Change in pitch angle is used to get most efficient power generation as small change in pitch angle can have large effect on power output. Positive pitch angle setting decreases the angle of incidence whereas negative pitch angle setting increases angle of incidence [9,16].

The relation between inflow angle ($\varphi$), angle of attack ($\alpha$), twist angle and pitch is given by eq. (6).

$$\varphi = \alpha + \theta \qquad (6)$$

Where, ($\theta$) is the combination of blade pitch angle ($\theta_p$) and twist angle ($\beta$)

$$\theta = \theta_p + \beta \qquad (7)$$

The area of the blade relative to rotor swept area is rotor solidity ($\sigma$) and has an optimum value for a given tip speed ratio. Decrease in rotor solidity results decrease in number of blades [5]. Equation (8) expresses rotor solidity in terms of chord ($C$), number of blades ($B$) and swept area.

$$\sigma = \frac{CB}{2\pi r} \tag{8}$$

The number of blades in modern horizontal wind turbine is variable. Most of them have two or three number of blades. The polar moment of inertia is independent of the azimuthal position of rotor and constant with respect to the yawing which ensures the smooth operation of the rotor in case of three-blade rotor. The difference in moment of inertia when the blades are at vertical and horizontal positions in case of two-blade turbine causes imbalance. It is also considered that two-bladed wind turbines are visually disturbing to the eye. Two-bladed wind turbine has less aerodynamic efficiency than a three bladed one, but it is comparatively cheaper. It is also more flexible in construction than three-bladed turbine. More than three blades are not preferred because of the additional cost associated. More number of blades for a constant solidity give rise to higher root stress and the blades are less stiff. [5,9].

The number of blades ($B$) can roughly be calculated by a relation expressed in eq. (9) [14].

$$B = \frac{80}{\lambda^2} \tag{9}$$

For design tip speed ratio five, the number of blades then is roughly three.

## 3. Scaled Rotor and Flow Similitude

For large size blades, the size of a conventional wind tunnel is inappropriate to conduct experiments. These performance evaluation tests are conducted by scaling down turbine in accordance to the available wind tunnel using similarity laws. Buckingham (or $\pi$) Theorem is used in dimensional analysis to obtain dimensionless parameters. These dimensionless parameters are used to make analogies between model and full scale turbine. Dimensionless numbers like rotor solidity ($\sigma$), Reynolds number ($R_e$), tip speed ratio ($\lambda$), Mach number ($M_a$), lift coefficient ($C_L$), drag coefficient ($C_D$) and ambient turbulent intensity (I) should be shared by two turbines to be in geometric, kinematic and dynamic similarity; i.e. all relevant parameters among these have same value for model and prototype [6,17].

### 3.1 Geometric Similarity

The model and full scale turbine should have same shape i.e. all linear dimensions should have same scale ratio. The ratio between two corresponding sections of model and prototype should be equal to a constant scale factor (say $h_g$). This constant factor is the geometric similarity coefficient and is expressed by eq. (10)

$$h_g = \frac{L_M}{L_P} = \frac{C_M}{C_P} = \frac{R_M}{R_P} \tag{10}$$

Where, $M$ – model, $P$ – prototype, $L$ – length of the blade, $C$ – chord, $R$ – radius of blade

Geometrically similar blades in similitude also requires both model and prototype to have same number of blades [6]. This can be expressed by equating rotor solidity of both model and prototype (see eq. (10) – (13)).

$$\sigma_M = \sigma_P \tag{11}$$

$$\frac{B_M C_M}{\pi R_M} = \frac{B_P C_P}{\pi R_P} \tag{12}$$

On rearranging eq. (12),

$$B_M = B_P \cdot \frac{C_P}{C_M} \cdot \frac{R_M}{R_P} \tag{13}$$

Equations (10), (11), (12), (13) lead to $B_M = B_P$ making it evident that model and prototype should have equal number of blades. Preservation of all angles, flow direction and orientation is also a requirement of geometric similarity [18].

The blade is scaled down by geometric similarity coefficient ($h_g$) equal to 0.06 resulting in model blade length 0.15 meters.

### 3.2 Kinematic Similarity

Kinematic similarity is obtained when the velocity triangle in the corresponding sections are geometrically similar. The velocities on model and prototype at homologous points are related by a constant factor and are in the same direction; i.e. flow should exhibit similar streamline. To obtain kinematic similarity geometric similarity is necessary [6].

Kinematic similarity coefficient ($h_c$) can be expressed as in eq. (14)

$$h_c = \frac{v_M}{v_P} \tag{14}$$

Equating Reynolds number,

$$R_{e_M} = R_{e_P} \tag{15}$$

$$\frac{\rho_P \, v_P \, C_P}{\mu_P} = \frac{\rho_M \, v_M \, C_M}{\mu_M} \tag{16}$$

Where, $\rho$ – air density, $\mu$ – dynamic viscosity
Equation (14), (15) and (16) lead to eq. (17).

$$\frac{v_M}{v_P} = \left(\frac{\rho_P}{\rho_M}\right) \cdot \left(\frac{\mu_M}{\mu_P}\right) \cdot \left(\frac{C_P}{C_M}\right) = h_c \tag{17}$$

For fluid at same temperature and pressure dynamic viscosity and density ratios equal 1, which leads to eq. (18)

$$\frac{v_M}{v_P} = \frac{C_P}{C_M} = \frac{1}{h_g} \tag{18}$$

This establishes the relation between geometric and kinematic similarity coefficient.

$$h_c = \frac{1}{h_g} \tag{19}$$

By equating the TSR of both model and prototype, the relation of rotational speed between them can be obtained. (see eq. 20)

$$\frac{\omega_M R_M}{v_M} = \frac{\omega_P R_P}{v_P} \tag{20}$$

Where, $\omega$ – angular velocity of turbine

$$\omega_M = \omega_P \cdot \frac{R_P}{R_M} \cdot \frac{v_M}{v_P} = \omega_P \cdot \frac{1}{h_g} \cdot h_c = \frac{\omega_P}{(h_g)^2} \tag{21}$$

$$N_M = \frac{N_P}{h_g^{\,2}} \tag{22}$$

Where, $N$ – rotational speed
These relations derived from Reynolds number matching, geometric scaling and TSR matching give impractical freestream and rotational velocities.

**3.3 Dynamic Similarity**

Model and prototype have dynamic similarity if the force acting at their corresponding points differ by a constant scaling factor. Coefficient of lift and drag, ($C_L$) and ($C_D$) at homologous sections should be equal which are functions of Reynolds number and angle of attack. Two turbines with same temperature and pressure will be in flow similitude when homologous sections have same twist and airfoil [6].
Scaling laws with and without Reynolds number matching can be referred to Burdett and Treuren [2].

## 4. Experimental Setup

**4.1 Physical Model Development**

The scaled wind turbine model was created using additive manufacturing methods using facilities available in Turbine Testing Lab (TTL) of Kathmandu University.
Additive manufacturing is a formal term for rapid prototyping or more popularly known as 3D printing. Additive manufacturing requires generation of 3D CAD (Computer Aided Design) model. The model is constructed by adding material in layer. Layer is a thin cross section of the of the CAD model. The use of additive manufacturing is beneficiary as parts with complex geometry can be manufactured without tooling. This reduces the manufacturing cost. Rapid prototyping is widely used in a variety of industries these days to create a prototype quickly before commercialization of the product [19-20].
3D printed wind turbines and the application of additive manufacturing in wind energy technology have huge potential. Engineers use rapid prototyping to print scaled model of bigger components and conduct experiments in controlled environment. Affordable 3D printing technology is very beneficial in development of small scale wind turbines. Reduced manufacturing cost can increase the accessibility for wind turbines and can be useful in rural electrification. – Basset et al. [21].
The turbine model in this study was printed using *Inspire D290* – a rapid prototype machine available at TTL (Turbine Testing Lab) of Kathmandu University. Individual blades were printed and assembled with a hub. The printer uses a production grade thermoplastic known as ABS (Acrylonitrile Butadiene Styrene) as its building and support material. It uses FDM (Fused Deposition Modeling) and the extrusion temperature for ABS is 225-230 ºC. The layer thickness of both build and support material is 0.15 millimeter to 0.4 millimeter.
The process of constructing a physical model using rapid prototype machine involves a number of procedural steps. It begins with the preparation of CAD model. SolidWorks was used to model wind turbine with coordinates extracted from QBlade. The model is then scaled (also considering printer build volume) and converted to STL (Stereo Lithography) file format. STL file formats are universally accepted standard in rapid prototyping industries to manufacture 3D models. The STL file is then transferred to rapid prototype machine and manipulated. Transformation, rotation and scaling can be done as per requirement. The model is then sliced into a number of layers. The thickness of layer determines the overall quality of printed product. The automated model construction begins after setting up the machine. Once the model building process completes, it is removed from the workbench and post-processing begins. As part of post processing the support material is detached from the constructed model [19, 22]. The printing process inside the rapid prototype machine and a complete printed single blade is shown in Fig. 3.

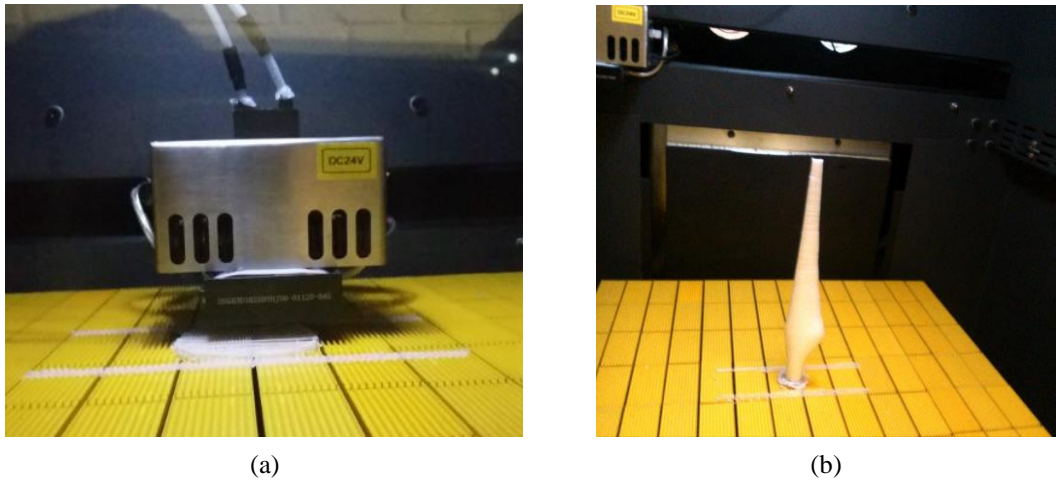

(a)            (b)

**Fig. 3** (a) 3D printing job of turbine model using *Inspire D290* and (b) 3D printed blade on workbench

**4.2 Turbine Stand and Motor**

A simple structure of nacelle was constructed to hold the motor and tower. The tower is designed (see Fig. 2.) to hold the turbine at a height where tower can capture most of the wind. The height of the tower is 0.18 m and is pinned to a base of $0.3 \times 0.3$ m$^2$. Stepper motor (EM-463) was used as a scaled generator as these motors have very good response to low rotational velocities. EM-463 stepper motor gives one volt per 100 rpm. The maximum operational current and power is 1 ampere and 48 watts respectively. The stepper motor's 5 mm thick shaft is coupled to the turbine's hub with a gear. The motor has a coil diameter of 0.28 mm and overall weight of motor is 54 grams. Variable electric resistance load was used to obtain optimum load value which corresponds to maximum power generation.

**4.3 Wind Tunnel**

Experiments were conducted in a low speed open circuit wind tunnel at Fluid Lab of Kathmandu University. The tunnel is 2.5 m long and has test cross section of $0.38 \times 0.38$ m$^2$. Figure 4. shows the schematic diagram of wind tunnel used in this experiment with front view of the test section.

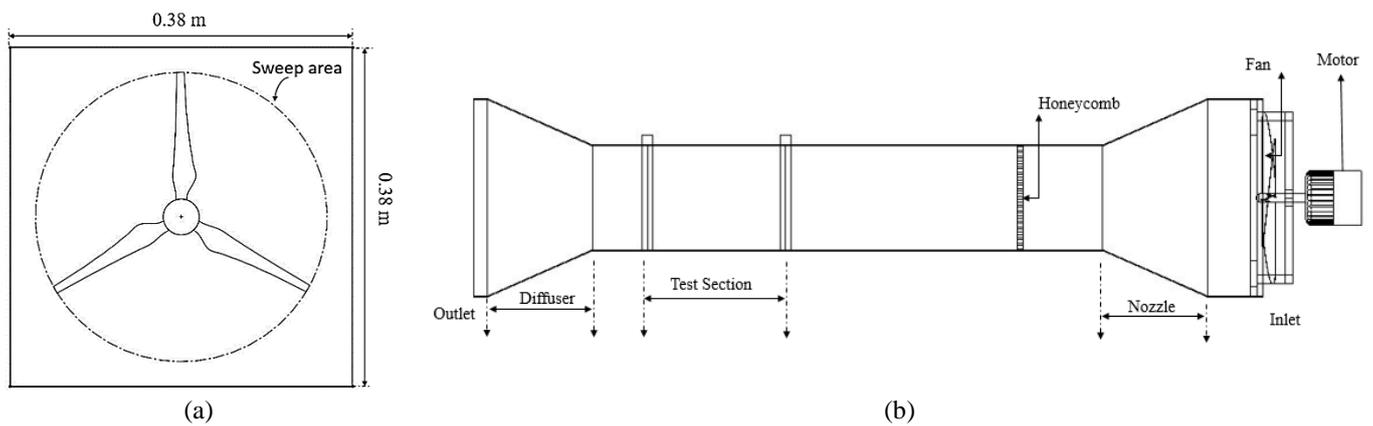

(a)            (b)

**Fig. 4** (a) Front view of test section and (b) Schematic diagram of wind tunnel

Experimental data necessary for aerodynamic analysis of wind turbine is obtained by conducting tests in this tunnel. The tunnel is a converging (nozzle) and diverging (diffuser) duct with a large axial fan whose rotational speed can be controlled and hence variable wind speed can be obtained. The upper limit of rotational speed of blower is 900 rpm and operates by DC motor supply. Honeycombs were used to reduce turbulence necessary to conduct tests.

Average cut-in speed of the wind turbine was 4.5 m/s and tests were conducted for wind speed ranging from 5 m/s to 7 m/s successively increasing by 0.5 m/s. Tests were not conducted for wind speed above 7 m/s as the turbine model could have attained higher rotational speed that would cause its structural failure. Four different turbine placements were chosen staring from test section, 100 mm, 200 mm, 300 mm and 400 mm (see Fig 9).

The power generated by the turbine model can be calculated by measuring different electrical quantities. Ohms law (see eq. 23)

can be used to calculate the electrical power generated when voltage and resistance are measured. Similarly, in order to find the extracted mechanical power of the rotor quantities like torque and angular speed are necessary.

$$P_E = \frac{V^2}{R_0}$$
(23)

Where, $P_E$ – Electric power, $V$ – voltage, $R_0$ – Resistance

## 5. Results and Discussions

### 5.1 Experimental Results

Wind tunnel tests were conducted for five different wind speeds. Four different placements were chosen in the test section. The first placement was 100 mm towards the outlet of the test section, progressively followed by 200 mm, 300mm, and 400mm (see Fig 9). Two different pitch settings, 0º and 5º were tested to observe the turbine's better response. In every case coefficient of performance was calculated and compared.

*Power, Efficiency and torque Observations*

The power production of turbine increases with increase in velocity (see Fig 5). Experimental power is measured by applying resistance load on the turbine. Observations show that the power produced is more in case of 5º pitch angle setting than in case of 0º. Among the four different placements maximum power production is obtained at 400 mm. In case of 0º pitch angle setting the power production trend follows the same pattern but with less magnitude.

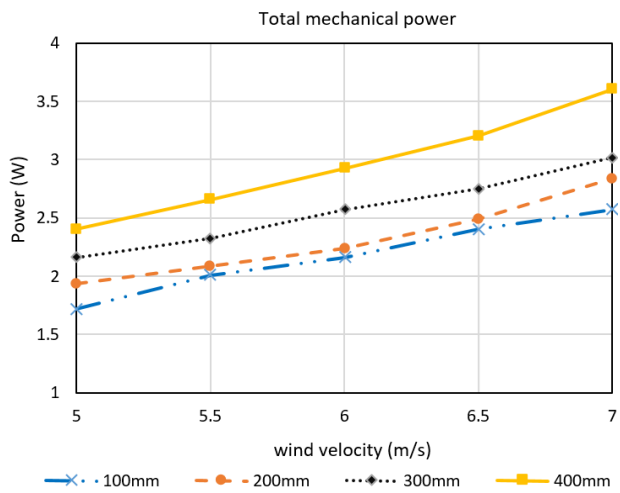

**Fig. 5** Experimental power distribution at all test placements, 5º pitch setting

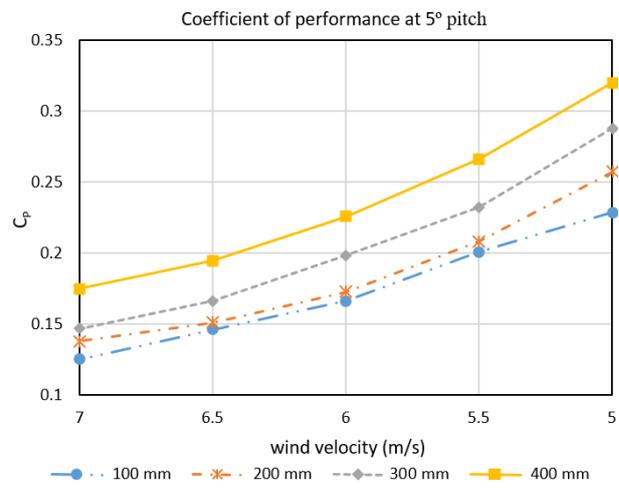

**Fig. 6** Coefficient of power curve at all test placements, 5º pitch setting

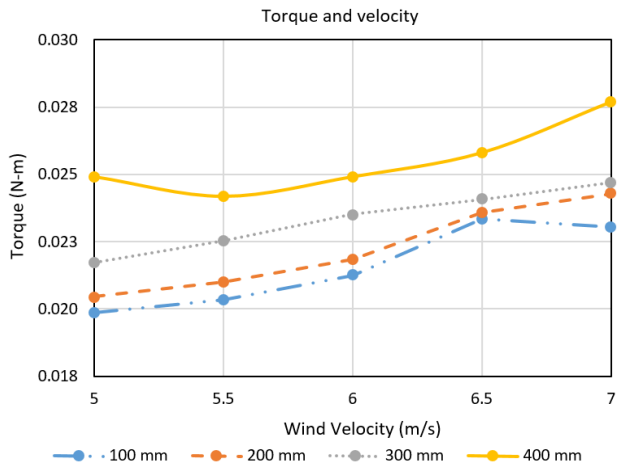

**Fig. 7** Torque versus wind velocity at different test section, 5º pitch angle

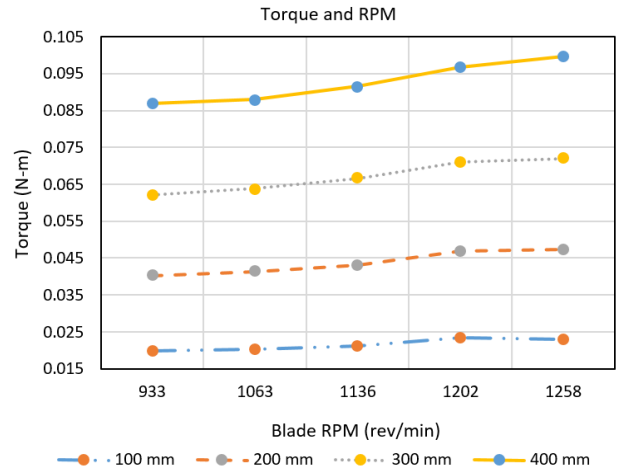

**Fig. 8** Torque versus rotor speed at different test placements, 5º pitch angle

The efficiency of the turbine is of major concern. The maximum coefficient of performance is obtained at design rated wind speed of 5 m/s as illustrated by the curves in Fig. 6. The best coefficient of performance is obtained at rated wind speed of 5 m/s, 5º pitch angle setting and 400 mm placement; and is equal to 0.32. Similarly, at 0º pitch angle setting the maximum coefficient of performance is 0.2 and the power curve traces similar pattern with lower numerical values.

Torque has a major role in mechanical power of the rotor. Low starting torques are favorable to conduct experiments. For constant power torque and angular speed compensate each other. Torque can be obtained using mathematical methods or by measurement. Opposite load either electrical or mechanical can be applied to the turbine and increased until it stops. The value of the load that stops the turbine can then be used to obtain torque. Test observation (see Fig. 7.) indicate low torque at lower wind speeds which gradually increase with increase with increasing wind speed and placements towards outlet of test section. The maximum value of torque obtained in 0.027 N-m at 5º pitch setting. Figure 8. shows the variation of torque with rotational speed of turbine rotor. It is evident that for all test placements torque has increments with increasing rotational speeds. The torque values are obtained for constant resistive load applied.

**5.2 Numerical Simulation**

Numerical simulations are carried out to observe the flow field and predict aerodynamic performance of wind turbine. Numerical results and experiments are used to validate each other. Numerical simulations were carried out using commercial software ANSYS CFX. *K-ε* turbulence model is used in the computational analysis. The fluid region is divided into two domains; stationary domain and rotating domain. The stationary domain represents the duct of wind tunnel whereas rotating domain is given rotational speed to represent the rotation of blade. Mesh grids on the blade were generated with growth rate 1 and inflation layer with layer quality 5 is used. It consists of 246,319 elements. In case of stationary domain, the number of elements is 1,387,110. Boundary conditions are chosen to replicate the wind tunnel test conditions. Reference pressure of 1 atmosphere is used and air density is taken as 1.25 kg/m$^3$ for air at 25 ºC. No slip wall condition is used at blade surface. Boundary conditions include inlet velocity, outlet pressure, wall. Fluid-Fluid interface is used between stationary and rotating domain with general connection of pitch angle 360º. The residual error is of order $10^{-4}$. Computational tests were conducted for pitch setting of 5º only and test placements 100mm, 200 mm, 300 mm, and 400 mm (see Fig. 9.)

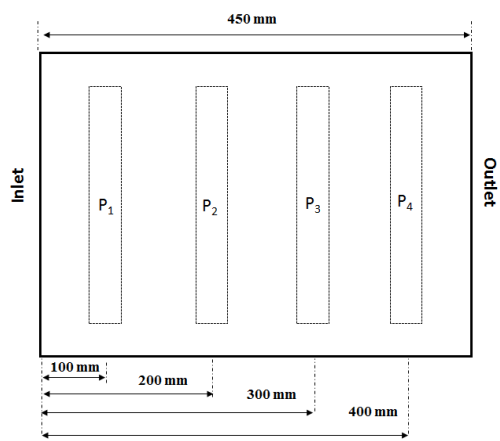

**Fig. 9** Test placements (P$_1$, P$_2$, P$_3$, P$_4$) computational domain (replicating test section placements in tunnel)

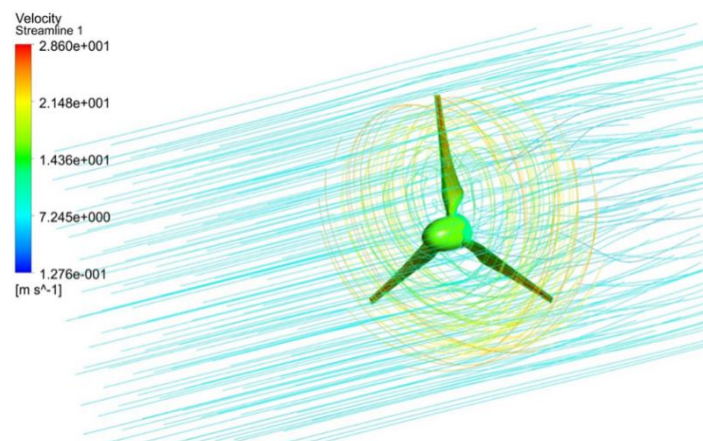

**Fig. 10** Velocity streamlines at 400 mm placement

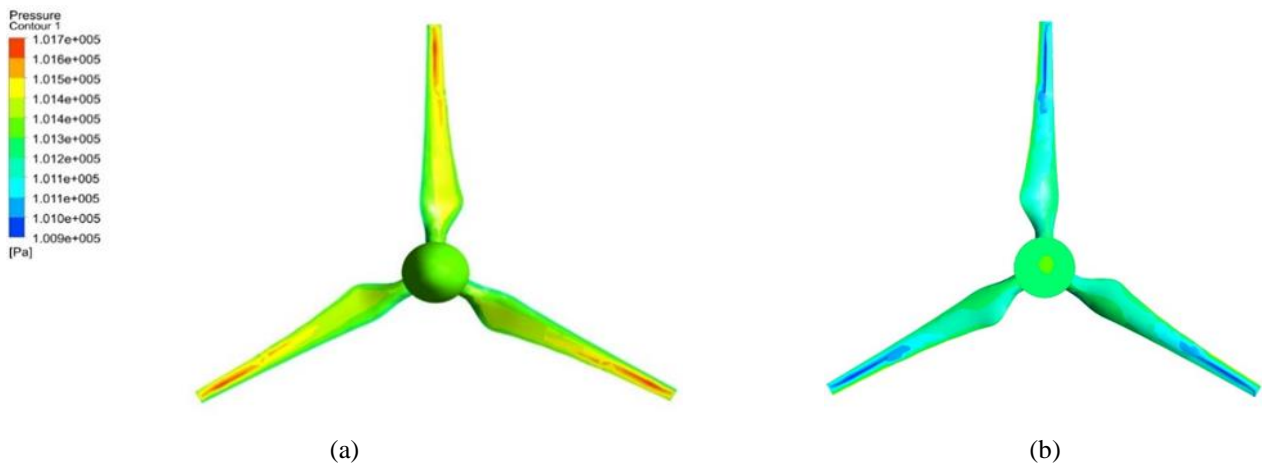

(a)          (b)

**Fig. 11** Pressure distribution on (a) pressure side and (b) suction side

From the streamlines obtained (see Fig. 10.) in rotating and stationary domain it is evident that the velocity is maximum at the tip of the blade which has NACA foil and it generates good lift. This produces 4.09 watts of power at 7 m/s wind speed. CFD calculations follow the same power distribution pattern as that of experimental.

## 5.3 Discussion

Power curves against wind velocity in both computational and experimental cases follow the same increasing pattern but differ quantitatively which can be seen in Fig. 12. BEM theory over predicts the performance of turbine than experimental and computational methods. Similarly, computational method's power prediction is also higher than the experimental method. On average at rated wind speed of 5 m/s, 5° pitch angle and 400 mm placement, average error in power prediction between computational and experimental method is 21.2 %. The discrepancy in experimental and computational torque maybe associated with generator inefficiencies like coil loss, eccentricity in shaft etc. These factors are also responsible for quantitative difference in power and efficiency given by both methods. Figure 7. illustrates the incensement of torque with wind velocity at the particular (optimum) resistive load applied at all four test placements.

One of the objectives of the project was to test the response of the turbine at different pitch angles and find the angle at which the turbine responded better. Tests for two different pitch angles were conducted. Experimental results illustrated the better performance of turbine at 5° pitch angle setting than 0°. The pattern of power production in case of both angles agree reasonably well against wind velocity. Power curve of Fig. 13 compares the performance of turbine at both pitch angles. The blades pitch angle in combination with blade twist changes the inflow angle (see eq. (6), (7)) and can have both positive and negative effect on turbine's performance.

Coefficient of performance obtained by QBlade is 0.5. Among all forty experimental test cases the maximum $C_P$ obtained is 0.32. In case of computational method, $C_P$ attains a maximum value of 0.36. These values are satisfactory compared to upper limit given by Betz i.e. 0.59. These maximum values of $C_P$ are obtained at rated wind speed of 5 m/s (see Fig. 14.).

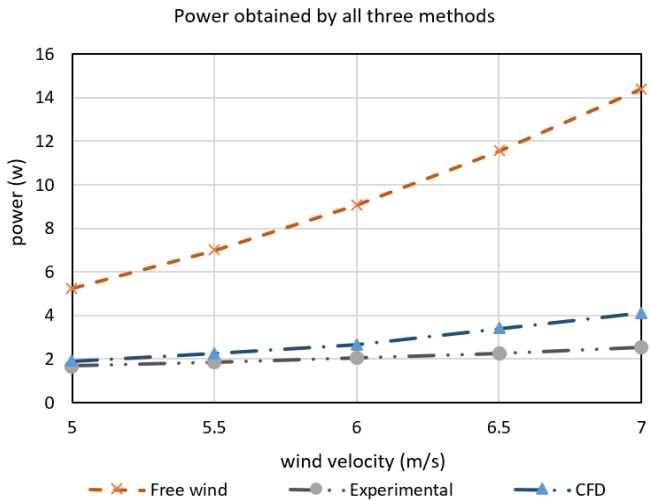

**Fig. 12** Theoretical, experimental (electric) and computational power curve

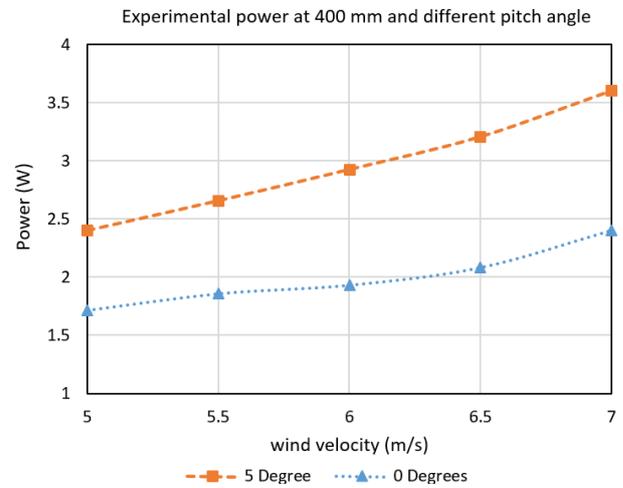

**Fig. 13** Power (mechanical) curve at different pitch setting, 400 mm placement

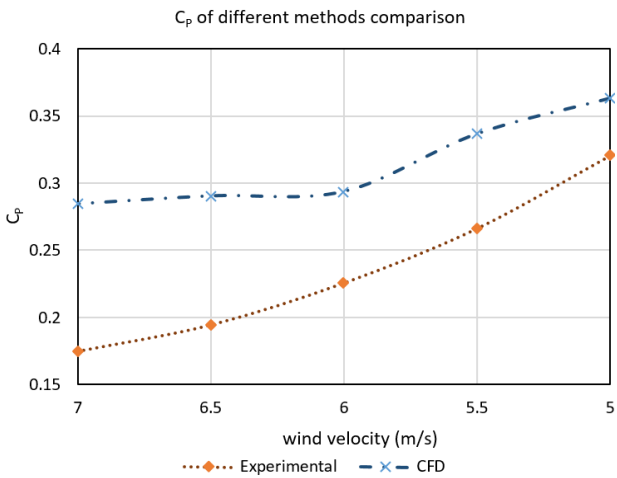

**Fig. 14** Coefficient of power curve for experimental and computational method

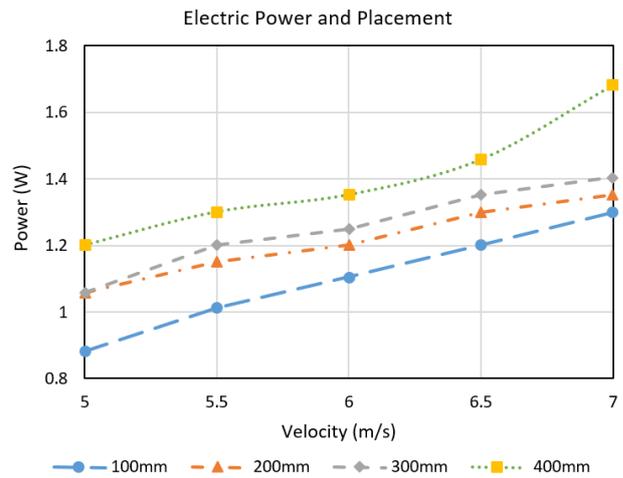

**Fig. 15** Total electric power at all test placements 0° pitch angle setting

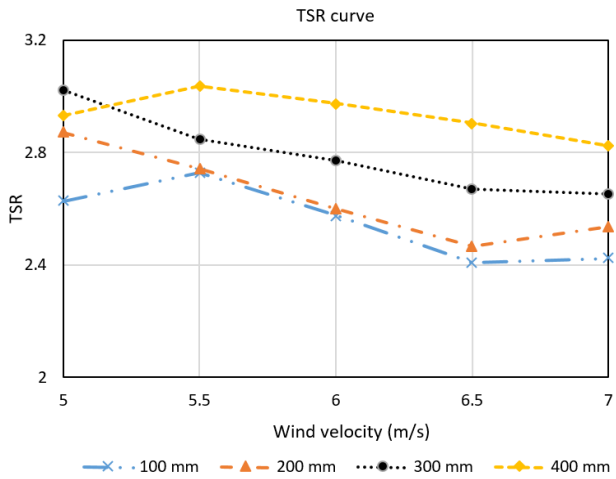 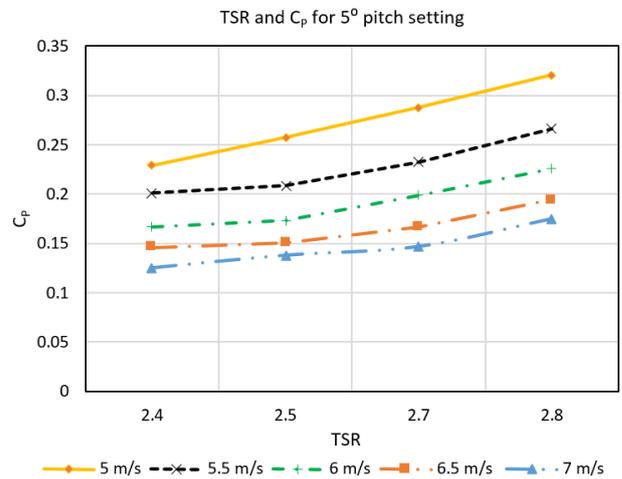

**Fig. 16** TSR distribution curve, 5º pitch setting    **Fig. 17** TSR versus power coefficient curve at all test speeds

For tests conducted at four different placements in the test section, maximum power is obtained at placement 400 mm towards the outlet of test section. The maximum electric power obtained at this placement, wind speed 7 m/s and pitch angle 5º is 2.5 watt. Similarly, at 0º pitch angle setting and same wind speed maximum power obtained is 1.68 watt. Similar trend is exhibited by computational results. The power obtained at rated wind speed 5 m/s and maximum test speed 7 m/s by computational method at 400 mm placement is 1.90 watt and 4.09 watt respectively. There is increase in power on moving from test placements P1 to P4 respectively. Total electric power obtained is less than mechanical power because of losses in motor coil and motor inefficiencies. Figure 15. is a power curve obtained for 0º pitch angle setting which also illustrates the increase in power production on moving test placement from 100 mm to 200 mm; maximum power is obtained at 400 mm.

The tip speed ratio for this tri-blade turbine design is 5. Experimental tip speed ratio is calculated using rotational speed of the bade. Coefficient of performance increases with increase in TSR at first and reaches a maximum value. On further increasing TSR, $C_P$ decreases. Figure 16. shows the distribution of TSR at all wind speeds considered in the test and for all test placements keeping the pitch angle setting as 5º. The maximum value of TSR obtained in experiment is 3. The relation between TSR and power coefficient for different test speeds at 5º pitch angle setting is shown in Fig. 17. For each wind speed chosen for test, the value of power coefficient varies over a range. Experimental value of TSR is smaller than the design value. Al-Abadi et al. [23] mention that parameters like rotors and motor sizes, resistive load govern experimental resulting in small range of TSR. Hence, complete behavior of power coefficient of a turbine in relation to TSR cannot be studied which is also necessary to predict turbine's performance.

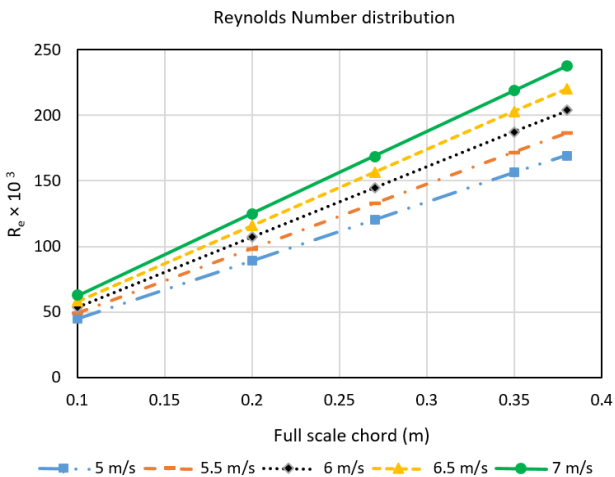 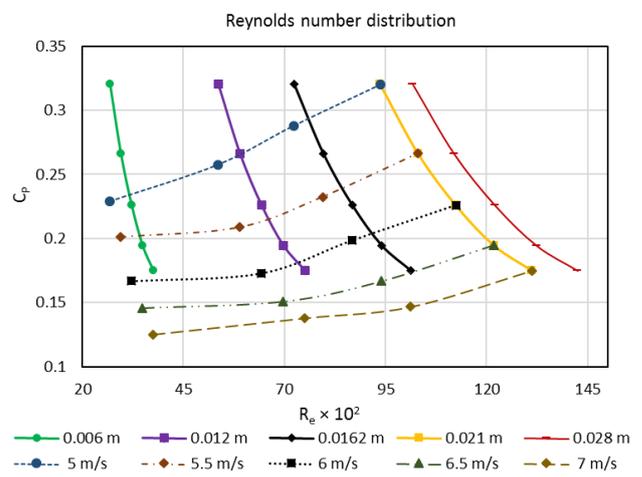

**Fig. 18** Reynolds Number distribution of full scale blade    **Fig. 19** Reynolds Number versus power coefficient of scaled blade at different chord length and wind velocity

There is significant effect and influence of surface roughness on the performance of wind turbine. The model was constructed by using 3D printing technology. So, the model has inherent roughness from printing process. Flow transition and separation is very sensitive to surface roughness. Burdett and Treuren [2] also mention the effect of surface roughness on flow separation. Roughness aids in transition from laminar to turbulent flow and reduces flow separation. The flow stays attached due to roughness and turbine performance increases. But the flow separation also depends on Reynolds number. The combination of Reynolds number and surface roughness have influence in turbine's performance. Figure 18 shows the distribution of Reynolds number with full scale chord length. Reynolds number experienced by turbine ranges from 44,642 to 237,500 for full scale blade and from 2,678 to 14,250 for scaled model. Figure 19. shows the influence of Reynolds number of on power coefficient at five different chord lengths and wind velocities. The curve shows the direct influence of Reynolds number on power coefficient at different wind velocities.

## 6. Conclusion

The performance of a HAWT blade designed using QBlade was studied using both computational and experimental methods. 2.5 m long blade was scaled to 0.15 m and a physical model of turbine was prepared using rapid prototype technology. Experiments were conducted in an open circuit wind tunnel and computational studies were made using ANSYS CFX. For experimental tests conducted at different placements in the test section the performance of the blade was comparatively better at 400 mm. There was a gradual decrease in performance on moving from test placement P4 to P3, P3 to P2 and P2 to P1 (i.e. from the inlet side of test section towards the outlet, see Fig. 9). The blade was also tested for two different pitch angle settings and its response was better at $5^0$ pitch angle than at $0^0$. Out of forty different test cases the maximum power coefficient obtained from experiment is 0.32 with pitch angle setting $5^o$ and 400 mm placement. Similarly, from computational method power coefficient with a maximum value of 0.36 was obtained at same test configuration. The performance prediction of computational method is slightly higher than the results obtained from experiments. Since model was constructed using 3D printing technology, it has surface roughness adequate to influence the turbines performance at the range of Reynolds number it operated. The Reynolds number was below 500,000 and results show its influence on turbines performance.

## Acknowledgments

The research was conducted at Kathmandu University with the support of Turbine Testing Lab. The authors would like to acknowledge the valuable contributions of everyone associated with the project.

## Nomenclature

| | | | | |
|---|---|---|---|---|
| $A$ | Sweep area [m$^2$] | | $\alpha$ | Angle of attack [$^o$] |
| $B$ | Number of blades | | $\beta$ | Twist angle [$^o$] |
| $C$ | Chord length [m] | | $\theta_p$ | Pitch angle [$^o$] |
| $C_D$ | Coefficient of drag | | $\lambda$ | Tip speed ratio |
| $C_L$ | Coefficient of lift | | $\mu$ | Dynamic viscosity |
| $C_P$ | Coefficient of lift | | $\varphi$ | Angle of relative wind [$^o$] |
| $h_c$ | Kinematic similarity coefficient | | $\rho$ | Air density [kg/m$^3$] |
| $h_g$ | Geometric similarity coefficient | | $\sigma$ | Rotor Solidity |
| $I$ | Turbulent intensity | | $\omega$ | Angular speed |
| $L$ | Length of blade | | | |
| $M_a$ | Mach number | | **Subscript** | |
| $N$ | Rotational speed | | | |
| $P$ | Wind power [W] | | $M$ | Model |
| $P_R$ | Rotor output [W] | | $P$ | Prototype |
| $P_E$ | Electric power [W] | | **Abbreviations** | |
| $r$ | Local radius of blade [m] | | | |
| $R$ | Radius of blade [m] | | ABS | Acrylonitrile Butadiene Styrene |
| $R_e$ | Reynolds number | | BEM | Blade Element Momentum |
| $R_0$ | Electric resistance (Ω) | | CAD | Computer Aided Design |
| $T_N$ | Thrust [N] | | CFD | Computational Fluid Dynamics |
| $T_Q$ | Torque [N-m] | | FDM | Fused Deposition Modeling |
| $v$ | Wind speed [m/s] | | HAWT | Horizontal Axis Wind Turbine |
| $V$ | Voltage | | TSR | Tip Speed Ratio |